\documentclass{sig-alternate-2013}

\usepackage{shortcuts,graphicx,caption,subcaption}
\usepackage{hyperref,soul}

\permission{Copyright is held by the International World Wide Web Conference Committee (IW3C2). IW3C2 reserves the right to provide a hyperlink to the author's site if the Material is used in electronic media.}
\conferenceinfo{WWW'14 Companion,}{April 7--11, 2014, Seoul, Korea.} 
\copyrightetc{ACM \the\acmcopyr}
\crdata{978-1-4503-2745-9/14/04. \\
http://dx.doi.org/10.1145/2567948.2579233}

\begin{document}

\title{Predicting Crowd Behavior with Big Public Data}

%
%
%
%
%

\numberofauthors{1} 
%
\author{
\alignauthor
Nathan Kallus\\
       \affaddr{Massachusetts Institute of Technology}\\
       \affaddr{77 Massachusetts Ave E40-149}\\
       \affaddr{Cambridge, MA 02139}
       \email{kallus@mit.edu}
}

\date{20 August 2013}

\maketitle 
\begin{abstract}
With public information becoming widely accessible and shared on today's web, greater insights are possible into crowd actions by citizens and non-state actors such as large protests and cyber activism. We present efforts to predict the occurrence, specific timeframe, and location of such actions before they occur based on public data collected from over 300,000 open content web sources in 7 languages, from all over the world, ranging from mainstream news to government publications to blogs and social media. Using natural language processing, event information is extracted from content such as type of event, what entities are involved and in what role, sentiment and tone, and the occurrence time range of the event discussed. Statements made on Twitter about a future date from the time of posting prove particularly indicative. We consider in particular the case of the 2013 Egyptian coup d'\'etat. The study validates and quantifies the common intuition that data on social media (beyond mainstream news sources) are able to predict major events.
\end{abstract}


%

\section{Introduction}

The manifestation of crowd actions such as mass demonstrations often involves collective reinforcement of shared ideas. In today's online age, much of this public consciousness and comings together has a significant presence online where issues of concern are discussed and calls to arms are publicized. The Arab Spring is oft cited as an example of the new importance of online media in the formation of mass protests \cite{dynamics}. While the issue of whether mobilization occurs online is highly controversial, that nearly all crowd behavior in the internet-connected world has some presence online is not. So while it may be infeasible to predict an action developing in secret in a single person's mind, the ready accessibility to public information on the web that future crowds may now be reading and reacting to or members of which are now posting on social media can offer glimpses into the formation of this crowd and the action it may take.

News from mainstream sources from all over the world can now be accessed online and about 500 million tweets are posted on Twitter each day with this rate growing steadily \cite{telegraph}. Blogs and online forums have become a common m\-e\-d\-i\-u\-m for public discourse and many government publications are offered for free online. We here investigate the potential of this publicly available information online for predicting mass actions that are so significant that they garner wide mainstream attention from around the world. Because these are events perpetrated by human actions, they are in a way endogenous to the system, enabling prediction.

But while all this information is in theory public and accessible and could lead to important insights, gathering it all and making sense of it is a formidable task. We here use data collected by Recorded Future (\url{www.recordedfuture.com}). Scanning over 300,000 different open content web sources in 7 different languages and from all over the world, mentions of events---in the past, current, or said to occur in future---are continually extracted at a rate of approximately 50 extractions per second. Using natural language processing, the type of event, the entities involved and how, and the timeframe for the event's occurrence are resolved and made available for analysis. With such a large dataset of what is being said online ready to be processed by a computer program, the possibilities are infinite. For one, as shown herein, the gathering of crowds into a single action can often be seen through trends appearing in this data far in advance.

We here study the cases of mass protests and politically motivated cyber campaigns involving an entity of interest, such as a country, city, or organization. We use historical data of event mentions online, in particular forward-looking mentions of events yet to take place, to forecast the occurrence of these in the future. We can make predictions about particular timeframes in the future with high accuracy.

We find that the mass of publicly available information online has the power to unveil the future actions of crowds. Measuring the trends in sheer numbers we are here able to accurately predict protests in countries and in cities and large cyber campaigns by target and by perpetrator. After assembling our prediction mechanism for significant protests, we investigate the case of the 2013 Egyptian coup d'\'etat and how well we were able to foresee the protests surrounding it.

The ability to forecast these things has important ramifications. Countries and cities faced with a high likelihood of significant protest can prepare themselves and their citizens to avoid any unnecessary violence and damages. Companies with personnel and supply chain operations can ask their employees to stay at home and to remain apolitical and can attempt to safeguard their facilities in advance. Countries, companies, and organizations faced with possible cyber campaigns against them can beef up their cyber security in anticipation of attacks or even preemptively address the cause for the anger directed at them.

Recent work has used online media to glean insight into consumer behavior. In \cite{hplabs} the authors mine Twitter for insights into consumer demand with an application to forecasting movie earnings. In \cite{onlinechatter} the authors use blog chatter captured by IBM's WebFountain \cite{webfountain} to predict Amazon book sales. These works are similar to this one in that they employ very large data sets and observe trends in crowd behavior by huge volumes. Online web searches have been used to describe consumer behavior, most notably in \cite{googletrends} and \cite{websearch}, and to predict movements in the stock market in \cite{googleipos}.

In \cite{radinsky} the authors study correlations between singular events with occurrence defined by coverage in the New York Times. By studying when does one target event ensue another specified event sometime in the future, the authors discover truly novel correlations between events such as between natural disasters and disease outbreaks. Here we are interested in the power of much larger, more social, and more varied datasets in pointing out early trends in endogenous processes (actions by people discussed by people) that can help predict all occurrences of an event and pinning down \textit{when} they will happen, measuring performance with respect to each time window for prediction. One example of the importance of a varied dataset that includes both social media and news in Arabic is provided in the next section.

We here seek to study the predictive power of such web intelligence data and not simply the power of standard machine learning algorithms (e.g. random forests vs. SVMs). Therefore we present only the learning machine (random forest in the case of predicting protests) that performed best on the training data and compare it to a data-poor maximum-likelihood random-walk predictor that predicts for the future the situation today.  The data used in this study has been made available for download at \url{www.nathankallus.com/PredictingCrowdBehavior/}.

We first discuss what predictive signals we expect to find in order to motivate our constructions. We then review how mentions of events, entities, and times are extracted from the wide breadth of sources. Using this data we develop a predictive mechanism to predict significant protests in countries and in cities. We consider the case of the 2013 Egyptian coup d'\'etat and conclude. In an appendix we consider finding more general patterns in the data, motivating an application of the na\"ive Bayes classifier to high-dimensional sequence mining in massive datasets, which we use to forecast cyber attack campaigns by target or perpetrator.

\section{Predictive signals in public data}\label{predictivesignals}

We begin by exemplifying anecdotally the precursory signals that exist in public data for large protests. On Sunday June 9, 2013 a Beirut protest against Hezbollah's interference in Syria turned violent when clashes with Hezbollah supporters left one protester dead \cite{NYTimesJune9}. The story was widely reported on June 9 including in Western media, attracting more mainstream news attention than any protest event in Lebanon in over a year marking it as a \textit{significant protest}.

But not only were there signs that the protest would occur before it did, there were signs it may be large and it may turn violent. The day before, Algerian news source Ennahar published an article with the headline ``Lebanese faction organizes two demonstrations tomorrow rejecting the participation of Hezbollah in the fighting in Syria'' (translated from Arabic using Google Translate). There was little other preliminary mainstream coverage and no coverage (to our knowledge) appeared in mainstream media outside of the Middle-East-North-Africa (MENA) region or in any language other than Arabic. Moreover, without further context there would be little evidence to believe that this protest, if it occurs at all, would become large enough or violent enough to garner mainstream attention from around the world.

\begin{figure}[b!]\vspace{-10pt}
\centering
\includegraphics[width=.45\textwidth]{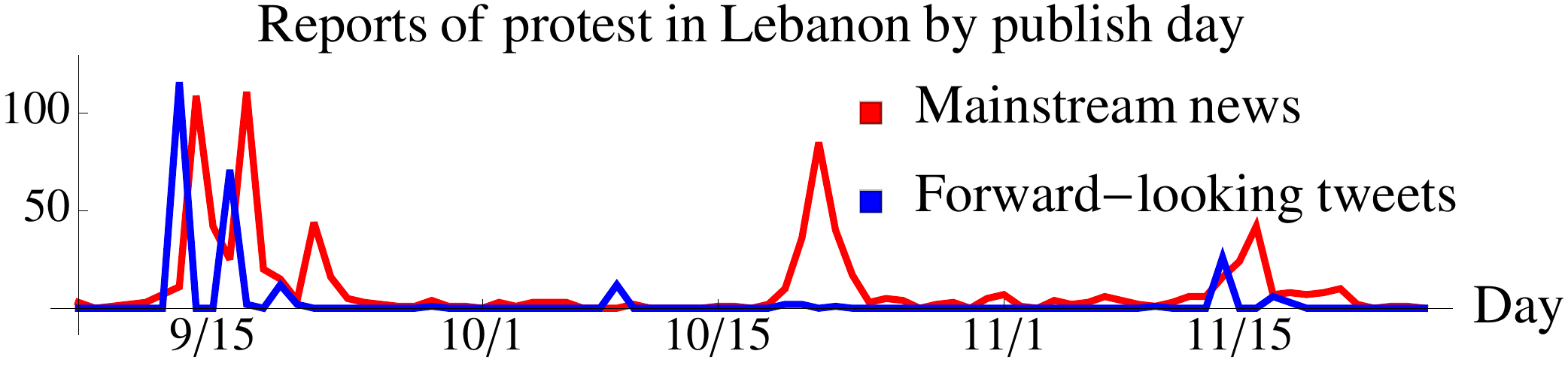}
\caption{Number of same-day news reports of protest in Lebanon (red) and tweets mentioning protests said to occur over the next three days (blue). Red spikes often follow blue spikes or otherwise have a recognizable convex ramp-up.}
\label{lebpatterns}\end{figure}

However, already by June 5, four days earlier, there were many Twitter messages calling people to protest on Sunday, saying ``Say no to \#WarCrimes and demonstrate against \#Hezbollah fighting in \#Qusayr on June 9 at 12 PM in Downtown \#Beirut'' and ``Protest against Hezbollah being in \#Qusair next Sunday in Beirut.'' In addition, discussion around protests in Lebanon has included particularly violent words in days prior. A June 6 article in TheBlaze.com reported, ``Fatwa Calls For Suicide Attacks Against Hezbollah,'' and a June 4 article in the pan-Arabian news portal Al Bawaba reported that, ``Since the revolt in Syria, the security situation in Lebanon has deteriorated.'' A May 23 article in The Huffington Post mentioned that, ``The revolt in Syria has exacerbated tensions in Lebanon, which ... remains deeply divided.'' Within this wider context, understood through the lens of web-accessible public information such as mainstream reporting from around the world and social media, there was a significant likelihood that the protest would be large and turn violent. These patterns persist across time; see Figure \ref{lebpatterns}.

It is these predictive signals that we would like to mine from the web and employ in a learning machine in order to forecast significant protests and other crowd behavior. In this case, it is critical that we spread a wide net on many sources so to catch mentions in non-Western media and foreign-language Tweets along with mentions in media (such as Reuters) and languages (such as English) with a more global reach. 

\section{Event extraction and temporal analysis}

To quantify these signals we will look at time-stamped event-entity data. The data harvesting process extracts mentions of events from the plethora of documents continually gathered from the over 300,000 sources being monitored. An important aspect is that the event mentions are tagged with the time range in which the event is said to occur in the mention so that forward-looking statements, such as plans to protest, can be directly tied to a future time and place. Event extractions are done in Arabic, English, Farsi, French, Russian, Spanish, and Simplified and Traditional Chinese.

There are several elements in the event, entity, and time extraction process. For each document mined from the web, an ensemble of off-the-shelf natural language processing tools are used to extract tokens (lemma, root, stem, and part of speech) and entities. Entities extracted by each tool are then combined and resolved into a database of canonical entities for disambiguation. Documents are categorized by topic. Entities that are not agreed upon by the various tools and are far-fetched given the topic are rejected. Ontologies of structured entity relationships constructed from online s\-o\-u\-r\-c\-es (such as DBpedia) are used to guide filtering and provide a gazetteer for additional entity extraction.

Given the set of filtered entities, a statement extractor links these entities to events stated in the document. Events are again extracted using an ensemble method of off-the-shelf tools and a custom made tool that relies on the above tokenization. Each event from the various extraction sources is matched up with a particular text fragment from the document that best represents it. Next, $n$-grams in the document are matched against phrase lists organized by sentiment or tone and the fraction of these in the fragment is recorded.

All time statements made in the document are separately extracted. The tokenization of the document is parsed by a dependency grammar using the data-driven parser MaltParser \cite{MaltParser} to construct a dependency graph of the document. This is used to find time statements, both relative (e.g. ``next summer'') and absolute. Comprehending these in machine time is based on several contextual cues. Cultural and regional categorizations are extracted from the document to inform such things as relevant hemisphere for seasons, which is the first day of the week, standard date formats (month first or day first), and timezone. Moreover, contexts such as publishing date are noted. Using these, all time statements made in the document are converted to standardized time-stamps, with specificity varying from second to year. Event mentions are then matched up with the most relevant time statement to the event statement based on sentence dependency.

In all, events are marked by type of event, time range of event, entities involved, role of entities involved, entities mentioned, sentiment and tone, and origin. Some post-processing is done on the event-entity level to further improve quality based on special curated ontologies. For example, known hacker groups such as Anonymous constitute one such ontology and if mentioned in a cyber event but not clearly as the perpetrator these are assumed to be so nonetheless. Similarly, impossible events are rejected. For example, one ontology keeps track of death dates of people based primarily on information harvested from Wikipedia and assists in rejecting the event ``Marco Polo will travel to China in 2015'' as impossible because Marco Polo is dead.

The precision of event extraction is measured by workers on the Amazon Mechanical Turk (mturkers). To test the precision of time-stamping, for each language language-specific Human Intelligence Tasks (HITs) are constructed for a random sample. Each HIT consists of the text fragment, the extracted time-stamp, and the question ``is extracted time right or wrong?'' and is given to three mturkers. It is declared successful if at least two answer ``right.'' For example, the precision of time-stamping in both English and Spanish is measured at 93\%, in Arabic at 90\%, and in Simplified Chinese at 82\%. The precision of the event extraction is measured similarly by type. Protest events in English come in at 84\%. Malware threat events in English come in at 96\% and in Simplified Chinese at 90\%.

\section{Predicting significant protests}

\begin{table}[t!]\centering\scriptsize
\begin{tabular}{l|rr||l|rr}
  Country & All & Twitter & Country & All & Twitter \\\hline
 Afghanistan & 60918  & 27655 & Lebanon & 44153  & 23394 \\
 Bahrain & 246136  & 177310 & Libya & 162721  & 69437 \\
 Egypt & 944998  & 397105 & Nigeria & 70635  & 38700 \\
 Greece & 122416  & 70521 & Pakistan & 289643  & 213636 \\
 India & 491475  & 274027 & S. Arabia & 39556  & 13670 \\
 Indonesia & 34007  & 17120 & Sudan & 28680  & 13654 \\
 Iran & 118704  & 53962 & Syria & 212815  & 79577 \\
 Italy & 65569  & 43803 & Tunisia & 99000  & 27233 \\
 Jordan & 35396  & 19369 & Yemen & 70583  & 16712
\end{tabular}
\caption{Protest event mentions in the corpus.}\vspace{-15pt}
\label{table:sources}\end{table}

We now turn our attention to the use of this event data to the prediction of significant protests around the world. Our first forecasting question will revolve around predicting significant protests on the country level and considering that country alone. That is, a significant protest is one that receives much more same-day mainstream reporting than is usual for that country. So while most days a business with supply chain operations in Egypt operate under the usual volatile circumstances (since 2011) of Egypt---certainly more volatile than, say, Jordan and receiving much more attention for it---they are interested in receiving advance notice of protests that are going to be larger and more dangerous than the ordinary for Egypt. The same for another milieu. At the same time, we will use past patterns from other countries to inform the prediction mechanism when making a prediction about one country. In fact, the prediction mechanism will not be knowledgeable of the particular country in question but instead just the \textit{type} of country, quantified by cluster membership. In an appendix we also investigate the question of predicting protests that are significant relative to a common baseline and on the city level instead.

We restrict to a selection of 18 countries: Afghanistan, Bahrain, Egypt, Greece, India, Indonesia, Iran, Italy, Jordan, Lebanon, Libya, Nigeria, Pakistan, Saudi Arabia, Sudan, Syria, Tunisia, and Yemen. We will consider all protest event mentions in any of these countries being published any time between January 1, 2011 and July 10, 2013 as the event mention corpus. January 1, 2011 up to March 5, 2013 will serve for supervised training (and validation by cross-validation) and since March 6, 2013 up to July 10, 2013 will serve for test and performance scoring. Let
$$M_{cs}(i,j)=\text{\parbox{6cm}{Number of event mentions of protest in country $c$ taking place on day $j$ extracted from documents published on day $i$ from sources of type $s$}}$$
We tabulate some totals of these numbers over the whole event mention corpus in Table \ref{table:sources}. For example, nearly one million mentions of a protest event in Egypt occur in the data, over a third of a million on Twitter.


\begin{figure*}[t!]
\centering
\includegraphics[width=\textwidth]{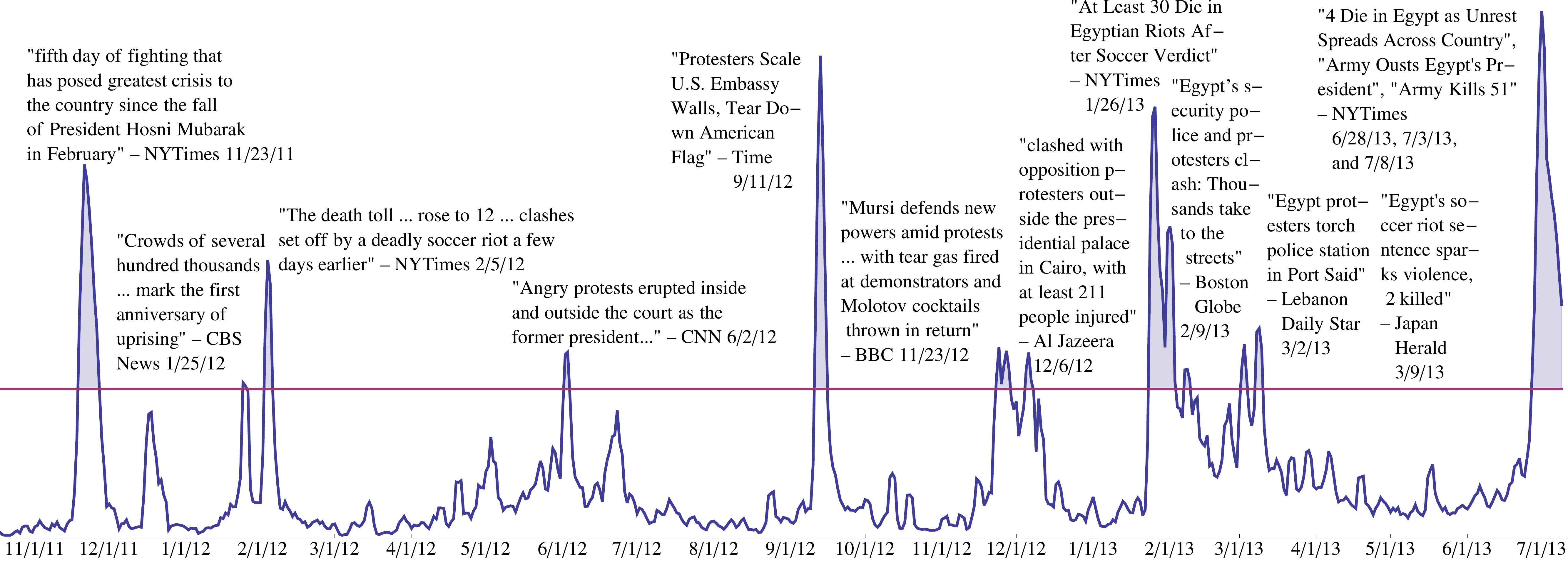}
\caption{Normalized count of mainstream reports $M''_c(i)$ in Egypt with annotations for stretches above $\theta=2.875$ (red line).
}\label{egypttimeline}\vspace{-10pt}
\end{figure*}

\subsection{The Ground Set}

For each country, the protests we are interested in forecasting are those that are significant enough to garner more-than-usual real-time coverage in mainstream reporting for the country. That is, there is a significant protest in country $c$ on day $i$ if $M_{c,\op{Mainstream}}(i,i)$ is higher than usual for country $c$ (in an appendix we consider an absolute scale for protest significance which naturally boosts predictability). Since new sources are being added daily by the hundreds or thousands to the Recorded Future source bank, there is a heterogeneous upward trend in the event mention data and what is more than usual in counts changes. To remove this trend we normalize the mention counts by the average volume in the trailing three months. That is, we let
\begin{align*}M'_{cs}&(i,i+k)=\\[-5pt]&\frac{M_{cs}(i,i+k)}{\frac{1}{\abs{\op{Countries}}\times90}\sum_{c'\in\op{Countries}}\sum_{j=i-90}^{i-1}M_{c's}(j,j+k)}\end{align*}
where $\op{Countries}$ is the 18 countries. Next we define the training-set average of same-day mainstream reporting
$$\overline{M'_{c}}=\frac{1}{\abs{\op{Train}}}\sum_{i\in\op{Train}}M'_{c,\op{Mainstream}}(i,i)$$
where $\op{Train}$ denotes the set of days in the training set.

Moreover, to smooth the data we consider a three-day moving average. Then we say, by definition, that a significant protest in country $c$ (and relative to country $c$) occurs during the days $i-1,i,i+1$ if $$M''_c(i)=\frac{1}{3}\sum_{j=i-1}^{i+1}\frac{M'_{c,\op{Mainstream}}(j,j)}{\overline{M'_{c}}}\;\geq\;\theta$$
is larger than a threshold $\theta$. The threshold is chosen so to select only significant protests. By inspecting the data's correspondence to the largest protests, we set $\theta=2.875$ (which is also nearly the $94\thh$ percentile of the standard exponential distribution). Overall across all countries considered, this resulted in 6\% of 3-day stretches to be labeled positive, distributed mostly evenly among the countries.

An example plot of $M''_c(i)$ for Egypt is shown in Figure \ref{egypttimeline} with annotations of top mainstream news describing the protests in each stretch of above-threshold days. Notable protest events provide the reader with anchor points are the 9/11 anniversary protests in 2012 (concurrent with the Benghazi attacks), the late-June protests leading up to the Egyptian coup d'\'etat, the riots set off by soccer-fan violence in early 2012, and the riots after 30 of the fans involved were sentenced to death which also coincided with riots connected to the anniversary of the revolution in early 2013.

\subsection{Scoring Protest Predictions}\label{setup}

We are interested in predicting on each day $i$ whether a significant protest will occur over the next three days $i+1,i+2,i+3$ based on information published on or before $i$. That is, on each day $i$ we wish to predict whether $M''_c(i+2)\geq\theta$ (which depends on days $i+1,i+2,i+3$).

We quantify the success of a predictive mechanism based on its balanced accuracy. Let $T_{ci}\in\{0,1\},~P_{ci}\in\{0,1\}$ respectively denote whether a significant protest occurs in country $c$ during the days $i+1,i+2,i+3$ and whether we predict there to be one. The true positive rate (TPR) is the fraction of positive instances ($T_{ci}=1$) correctly predicted to be positive ($P_{ci}=1$) and the true negative rate (TNR) is the fraction of negative instances predicted negative. The balanced accuracy (BAC) is the unweighted average of these: $\op{BAC}=\prns{\op{TPR}+\op{TNR}}/{2}$.
BAC, unlike the marginal accuracy, cannot be artificially inflated. Always predicting ``no protest'' without using any data will yield a high 94\% marginal accuracy but only 50\% balanced accuracy. In fact, a prediction without any relevant data will always yield a BAC of 50\% on average by statistical independence.

\subsection{The Features}

We now attempt to quantify the predictive signals we encountered anecdotally in Section \ref{predictivesignals}. These features will serve as the data based on which we make predictions.

In Section \ref{predictivesignals} we exemplified how the violence in language surrounding discussion of protest in a country can help set the context for the danger of a future protest to get out of hand. Each event mention is rated for violent language by the fraction of $n$-grams in the corresponding fragment (ignoring common words) that match a phrase list. Let
$$V_{cs}(i)=\text{\parbox{6cm}{Total violence rating of fragments associated with event mentions of protest in country $c$ extracted from documents published on day $i$ from sources of type $s$}}$$
Similarly to the normalization of event mentions due to the ever-growing source bank, we normalize this quantity as
$$V'_{cs}(i)=\frac{V_{cs}(i)}{\frac{1}{\abs{\op{Countries}}\times90}\sum_{c'\in\op{Countries}}\sum_{j=i-90}^{i-1}V_{c's}(j)}$$

In addition, forward-looking mentions in mainstream news and Twitter can help indicate whether a protest is planned and estimate how many might plan to attend. We have already defined $M_{cs}(i,i+k)$ which counts this data for $k\geq1$.

In order to facilitate trans-country training, we normalize these features with respect to the series we would like to predict, $M'_{c,\op{Mainstream}}(i,i)$. Similar to the normalization of\break $M''_c(i)$, we normalize these features by a per-country constant coefficient $\prns{\overline{M'_c}}^{-1}$.

For the purposes of trans-country training, we hierarchically cluster the countries using Ward's method \cite{ward} applied with the distance between two countries $c,c'$ equal to the Kolmogorov-Smirnov uniform distance between the distribution functions of the set of training values of $M''$ ignoring the time dimension. That is, $d(c,c')=$
\begin{align*}
\sup_{z\geq0}\abs{\frac{1}{\abs{\op{Train}}}\sum_{i\in\op{Train}}\prns{\mathbb I\braces{z\geq M''_c(i)}-\mathbb I\braces{z\geq M''_{c'}(i)}}}\end{align*}
This distance is also a non-parametric test statistic to test the the hypothesis that two samples were drawn from the same distribution. We construct $\lfloor2\sqrt{\abs{\op{Countries}}}\rfloor$ clusters using \textit{R} function hclust \cite{R}. We include as a feature the indicator unit vector of cluster membership of the country $c$ associated with the instance $(c,i)$. Thus the classifier does not know the particular country about which it is making a prediction, just its type as characterized by this clustering.

For each instance $(c,i)$ we also include as features the ten most recent days of same-day reporting on protest in $c$,
$$\frac{M'_{c,\op{Mainstream}}(i,i)}{\overline{M'_c}},\dots,\frac{M'_{c,\op{Mainstream}}(i-9,i-9)}{\overline{M'_c}}$$
along with the two most recent differences of these values. We also include the violence rating in recent mainstream reporting as the cumulative partial sums of the values
$$\frac{V'_{c,\op{Mainstream}}(i)}{\overline{M'_c}},\dots,\frac{V'_{c,\op{Mainstream}}(i-9)}{\overline{M'_c}}$$

Next we include the counts of mentions of protests said occur over the next three days, published either in mainstream news or in Twitter over the ten recent days. We incorporate this feature as the cumulative partial sums of the values
$$\frac{\sum_{k=1}^3M'_{cs}(i,i+k)}{\overline{M'_c}},\dots,\frac{\sum_{k=1}^3M'_{cs}(i-9,i+k)}{\overline{M'_c}}$$
for $s=\op{Mainstream}$ and $s=\op{Twitter}$.

When predicting farther into the future, about the three days starting with the $k\thh$ day from today, we push all indexes back by $k-1$ days, thus excluding any future data and maintaining the same overall length of the feature vector.

\begin{figure}[b!]\vspace{-10pt}
\centering
\includegraphics[width=.45\textwidth]{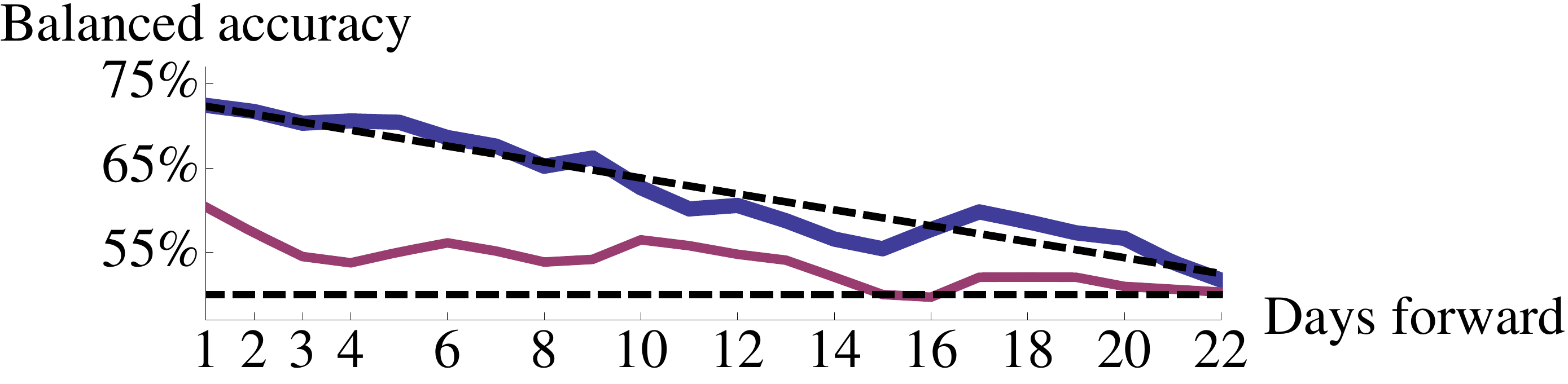}
\caption{BAC for predicting protests by distance into the future using the random forest and the full data-set (blue and trend in dashed black) and the data-poor predict-like-today heuristic (red).}
\label{fwd}\end{figure}
\begin{figure}[t!]
\centering
\begin{subfigure}{.24\textwidth}\includegraphics[width=\textwidth]{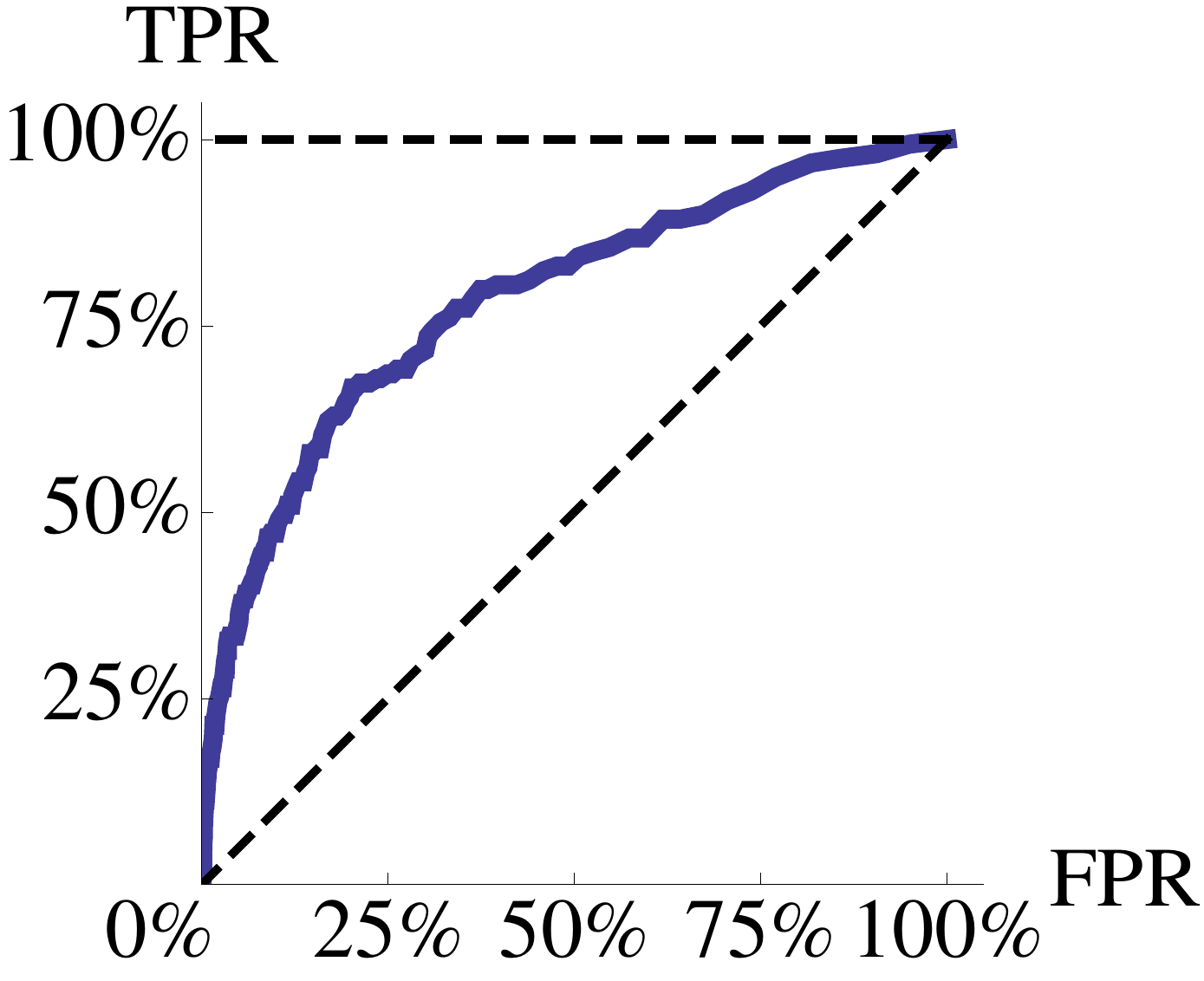}\end{subfigure}%
\hspace{-9pt}\begin{subfigure}{.24\textwidth}\includegraphics[width=\textwidth]{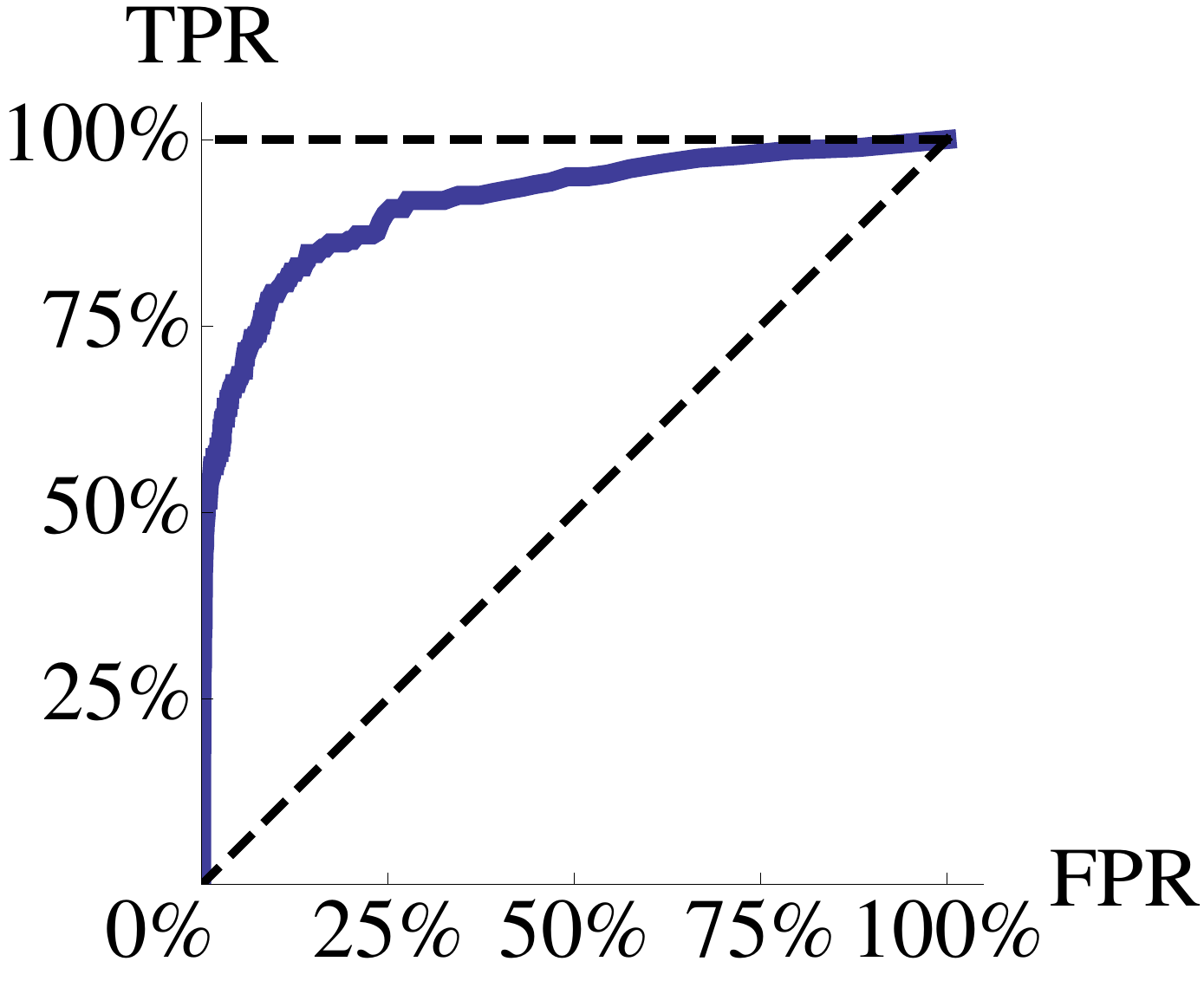}\end{subfigure}
\caption{Achievable TPR-FPR rates as we vary the classification threshold for (left) country-level predictions with a relative significance scale and (right) city-level predictions with an absolute significance scale.}
\label{roc}\vspace{-15pt}\end{figure}

\subsection{The Classifier}

For the prediction mechanism we employ a random forest classifier trained on all data up to March 5, 2013. We use the \textit{R} library randomForest \cite{Rrf}. The parameters are left at their default values as set in the library. For example, the defaults dictate that the forest have 500 trees each trained on $\lfloor\sqrt{\text{\#features}}\rfloor$ randomly chosen features.

We tune only the threshold required for a positive prediction. If the fraction of trees in the forest voting positive is at least this threshold then a positive prediction is made, otherwise negative. We tune this by four-fold cross validation over the training set using the BAC metric.

\subsection{Results}

We tested the trained random forest on test data March 6, 2013 to July 10, 2013. The results were TPR=\textbf{75.51\%} and TNR=\textbf{69.31\%}. BAC, the average of these two, is \textbf{72.41\%}, which constitutes a \textbf{44.8\%} reduction in balanced error from having no data. In comparison, predicting for the future the situation today, simulating the data-poor prediction possible when one nonetheless has information about today's situation (whether from being at the location or from news), has TPR=\textbf{27.04\%}, TNR=\textbf{93.74\%}, and BAC=\textbf{60.39\%}.

As we attempt to predict farther into the future our predictions become noisier and closer to the no data case as the earlier data has less bearing on the far future and there are fewer reports mentioning events to occur on the days in question. In Figure \ref{fwd} we plot the accuracy of making predictions farther into the future. For each $k\geq1$ we re-train the random forest with the pushed-back feature vectors.

As we vary the voting threshold, we can (monotonically) trade off true positives with true negatives. We plot the range of achievable such rates for our classifier in Figure \ref{roc} (left), using the false positive rate ($\op{FPR}=1-\op{TNR}$) as is common by convention. The area under the curve (AUC), here \textbf{78.0\%}, is the probability that a random positive instance has more positive votes than a random negative instance. The fraction of trees voting positive on an instance could well serve as a risk rating. By randomizing between two thresholds, we can achieve any weighted average of the rates. The rates achievable via randomization is the convex hull of the curve in Figure \ref{roc} and has \textbf{79.2\%} under it.

To show the possible breadth of this approach, we can also consider an alternative forecasting question where we instead wish to predict protests on the city level and we define significance on a global scale (see appendix for more details). That is, a protest is deemed significant if it garners unusually high same-day mainstream reporting relative to all cities considered. Using a global scale, the fraction of trees voting positive serves as an absolute risk rating that is comparable between cities. The predictability of positive events also increases because knowing the city alone is no longer statistically independent of there being a positive event (for example, 23\% of positive training instances are in Cairo). The further localized data also improves predictability. On the test data, the random forest yields TPR=\textbf{84.7\%}, TNR=\textbf{85.7\%}, BAC=\textbf{85.2\%}. The achievable rates as the voting threshold is varied are shown in Figure \ref{roc} (right) with AUC=\textbf{91.3\%} and \textbf{91.9\%} for the convex hull.

\subsection{The Case of the Egyptian Coup d'\'Etat}

\begin{figure}[t!]
\centering
\includegraphics[width=.47\textwidth]{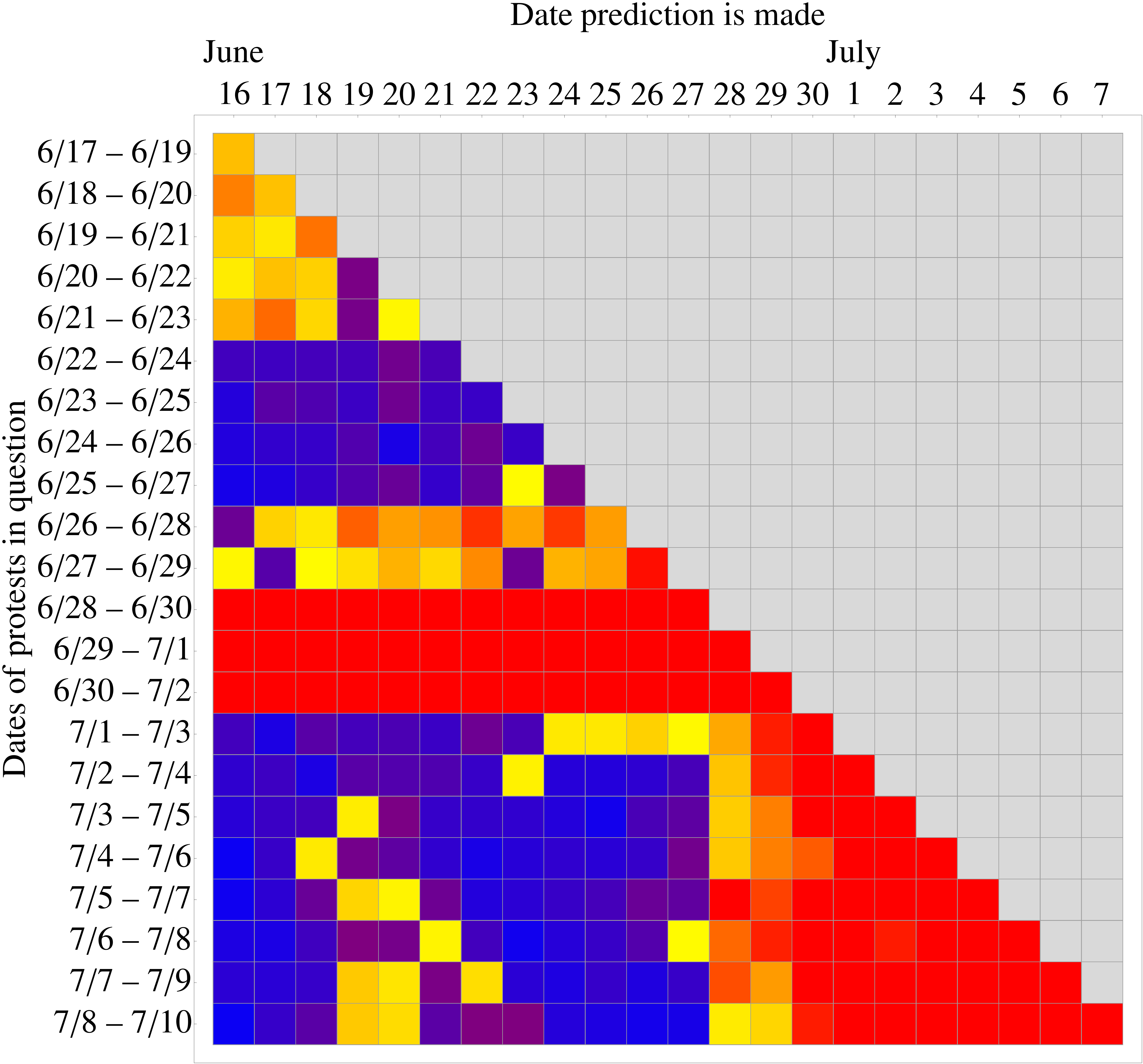}
\caption{Predictions of protests in Egypt around the time of the coup d'\'etat. Yellow to red mark positive predictions and blue to purple negative, with redder colors indicating more positive votes in the forest.}
\label{coup}\vspace{-15pt}\end{figure}

To exemplify the prediction mechanism we consider the case of the 2013 coup d'\'etat in Egypt and the protests surrounding it. Figure \ref{coup} depicts the predictions made on different days about different days in the relative future.

As can be seen, the days around June 30 were predicted positive with very high certainty for a long time prior, the date ranges in 6/28--7/2 being consistently predicted positive since June 6, three weeks beforehand (since June 16 onward shown in figure). Indeed, with a lot of discontent with and talk of demonstration against President Morsi's rule, many protests were anticipated for the weekend of June 30, the anniversary of Morsi's rise to the presidency. Even U.S. Secretary of State John Kerry made a statement in anticipation of protests asking for peaceful and responsible demonstration on, as he says, Saturday (June 29) and Sunday (June 30) \cite{kerry}. Therefore, those days were long predicted positive with high certainty. However, already on June 28 spontaneous ``warm-up'' protests burst in the streets \cite{warmup}. The first range to include June 28 was predicted with less certainty, especially from farther back, but was correctly predicted starting June 10 except on June 16 when it was mistakenly reported negative, just slightly below the threshold.

As we now know the protests around the anniversary indeed grew very large with many injured and dead in clashes with police and demonstrators from opposite camps. The protests did continue and on July 1 the Egyptian army issued an ultimatum to Morsi to resolve the protests within 48 hours or the army would intervene. On July 3 the Egyptian army led by General Abdul Fatah al-Sisi removed President Morsi from power. Protests intensified and many more people died. As seen in the figure, already on June 28 when the protests had only just started, before the anniversary and before any discussion of a possibile ultimatum or coup, the prediction mechanism had already correctly declared that significant protests will go on for the weeks to come.

\section{Conclusions}

We presented new findings about the power of massive online-accessible public data to predict crowd behavior. With much of public discourse having at least some presence online and usually more, the wide range of public data captured by our efforts offers unparalleled insight into the futures of countries, cities, and organizations as affected by mass demonstrations and cyber campaigns (see appendices). The evidence presented validates and quantifies the common intuition that data on social media (beyond mainstream news sources) are able to predict major events. The scope and breadth of the data offered glimpses into tweets in foreign languages and news in far places. The confluence of all this information showed trends to come far in the future. We are excited to make the data used in this study public and invite further exploration into its predictive abilities.

\bibliographystyle{abbrv}
{\small\bibliography{predictivepower}}  
%
%

\pagebreak
\section{Appendix: Predicting Protests on the City Level}

This appendix provides further detail on the prediction of protests on the city level and on a common scale of protest significance. From within the countries considered previously, we choose the top cities by number of mentions of protest events. These 37 cities are Jalalabad, Kabul, and Kandahar in Afghanistan; Manama in Bahrain; Alexandria, Cairo, Port Said, and Tanta in Egypt; Athens in Greece; Hyderabad, Mumbai, and New Delhi in India; Jakarta in Indonesia; Tehran in Iran; Milan and Rome in Italy; Amman in Jordan; Beirut and Sidon in Lebanon; Benghazi and Tripoli in Libya; Abuja and Lagos in Nigeria; Islamabad, Karachi, Lahore, Peshawar, and Quetta in Pakistan; Qatif and Riyadh in Saudi Arabia; Khartoum in Sudan; Aleppo, Damascus, Deraa, Hama, and Homs in Syria; Tunis in Tunisia; and Sana'a and Taiz in Yemen. We consider the same time range and the same train-test split.

We define $M'$ and $V'$ as before but for these cities. Since we are interested in an absolute level of significance we no longer normalize with respect to the entity, only with respect to the cross-entity average trailing volume. Thus, while overall still only 6\% of instances are labeled positives, Cairo takes up 23\% of positive training instances and 40\% of positive test instances, while Khartoum takes up 1\% of positive train instances and has no positive test instances.

Nonetheless, as before, cities are clustered according to their set of $M'$ training values and the classifier is not knowledgeable of the particular city in question, just its cluster membership. We use as features the unnormalized violence rating of past ten days of mainstream reporting about the city, same-day mainstream reporting level of past ten days, and the forward-looking mainstream reporting and Twitter discussion of past ten days. In addition, we include the unnormalized features of the containing country.

A random forest classifier is trained with the voting fraction threshold tuned by four-fold cross-validation to maximize balanced accuracy. Any other parameters were set to their defaults as before. Testing on March 6, 2013 to July 10, 2013, we get a true positive rate of \textbf{84.7\%} and a true negative rate of \textbf{85.7\%} yielding a balanced accuracy of \textbf{85.2\%}. The achievable rates as the voting fraction threshold is varied are shown in Figure \ref{roc} (right). The area under the ROC curve (AUC) is \textbf{91.3\%} and the area under its convex hull is \textbf{91.9\%}.

\section{Appendix: Predicting cyb\-er at\-t\-acks by sequence mining with na\-\"i\-ve Bayes}

In this appendix we expand our scope and consider all event types recorded. There are 112 distinct event types in the data ranging from protest to company acquisition to cyber attack to music album release to voting result. We wish to be able to predict unusually high numbers of mentions of a particular type of event involving a particular entity. Here we focus on cyber attacks. However, there are varying levels of ``clumpiness'' for the many classes of events and entities in terms of how and for how long a real-world event is discussed online. In addition, it is often hard to hypothesize \textit{a priori} what predictive signals may exist. Therefore, in order to tackle this forecasting problem we would need to spread a wider net and consider all event interactions and at the same time allow for more smoothing.

We will therefore consider events on the week level for a given entity $n^*$ (which could be a country, a person, an organization, a product, among many other things) and use events mentioning that entity to forecast the level of mentions of an event type of interest involving that entity next week. Let
$$M_{n^*es}(i,j)=\text{\parbox{6cm}{Number of event mentions of type $e$ involving entity $n^*$ taking place on week $j$ extracted from documents published on week $i$ from sources of type $s$}}$$
Here we will consider source types Any, Mainstream, Social Media, and Blog. As before, we normalize this number with respect to the total event mention volume in the past 12 weeks (approximately three months, as before) in order to de-trend it as follows
$$M'_{n^*es}(i,i+k)=\frac{M_{n^*es}(i,i+k)}{\sum_{e'\in\op{EventTypes}}\sum_{j=i-12}^{i-1}M_{n^*e's}(i,i+k)}$$
This is the data we will feed to our prediction algorithm. We will consider both the mentioning over the past weeks of events taking place in that same week as well as any forward-looking mentions on a past week of events to take place next week, the week in question. As before, we will also consider the case where we must predict farther into the future, about the week after next or the one after that etc.

We will consider data starting from the first week of 2011 and up to the last week of July 2013. We will test our mechanism on April 2012 onward, training on the trailing two years (as available). Any cross-validation is done on 2011 up to March 2012.

\subsection{The Ground Set}

Along with an entity of interest $n^*$, let us fix an event type of interest $e^*$ and a source type of interest $s^*$. Because we believe our data is particularly useful in predicting crowd behavior we will choose $e^*$ accordingly. Here we will be interested in predicting politically motivated cyber campaigns so we fix $e^*=\op{CyberAttack}$. We label as positive weeks that included cyber attack campaigns that were so impactful to generate wide attention all over with same-week mentions of cyber attack events. Therefore we fix $s^*=\op{Any}$. We will consider $n^*$ that are both country target entities (such as Israel) and hacktivist attacker entities (such as Anonymous).

We also fix a threshold $\theta$ and we will wish to predict on week $i$ whether
$$M'_{n^*e^*s^*}(i+,i+)\geq\theta$$
We fix $\theta$ so that 15\% of weeks are positive.

As before, we will use balanced accuracy to score our predictive mechanism and to tune parameters by cross-validation.

\subsection{High-Dimensional Sequence Mining with Na\"ive Bayes}

Let $T_i=1$ denote the positivity of the prediction instance on week $i$
$$T_i=1\;:\;M'_{n^*e^*s^*}(i+1,i+1)\geq\theta$$
We seek to estimate the conditional probability density conditioned on the past $\ell$ weeks
\begin{equation}
\label{density}\mathbb P\prns{T_i=t\;\left|\;\begin{array}{l}M'_{n^*es}(i-k,i-k),\,M'_{n^*es}(i-k,i+1)\\\text{ for }e\in\op{EventTypes},\\\phantom{\text{ for }}s\in\op{SourceTypes},\\\phantom{\text{ for }}k=0,\dots,\ell-1\end{array}\right.}\end{equation}
for $t=0$ or $1$. We use $\ell=5$ here. 

That estimate this, we seek to find the patterns of $\ell$ event sequences that end with our target event. Sequence mining is the discovery of commonly repeating sequences in a string over an alphabet $\Sigma$. In bioinformatics, sequence mining is applied to DNA sequences ($\Sigma=\{A,C,G,T\}$) and to amino acid sequences constituting a protein ($\abs{\Sigma}=20$) to find common sequences of some length. For longer strings the frequency of appearing in nature is highly concentrated.

We first bin the values of $M'_{n^*es}(i,i+k)$ into quartiles of their marginal distribution over the training data. The resulting alphabet $\Sigma$ is of size $4^{2\times\ell\times\abs{\op{SourceTypes}}\times\abs{\op{EventTypes}}}$, much larger than the training data set so that the probability function is underspecified (high-dimensional setting). At the same time, the amount of information in the training data is also quite massive. So we require a method that can smooth the density to avoid overfitting and, at the same time, tractable over a large data set.

One solution to this problem is to apply what is known as the na\"ive assumption to our likelihood function \eqref{density}. Let $F,F'\in\{1,2,3,4\}$ (indicating the binned quartile) be any two different features of the past in the conditioning in \eqref{density}. Then we make the following assumption:
\begin{equation*}
\text{Conditioned on $T_i$, $F$ and $F'$ are statistically independent.}
\end{equation*}
This is of course very different from assuming marginal independence. For example under the assumption, discussion of protests or of military actions in Israel in mainstream news could very often coincide with a discussion on Twitter of a planned future cyber attack against Israel because a possible future cyber attack is often a response to the former; but given that a cyber attack against Israel does occur next week, we assume the former two must occur independently from one another.

By Bayes' theorem we may decompose the probability function to a product of $\mathbb P\prns{T_i=t}$ and the conditional probabilities of $M'_{n^*es}(i-k,i-k)$ and $M'_{n^*es}(i-k,i+1)$ given $T_i=t$. Estimating instead the marginal distribution of $T$ and the conditional distributions of $M'$ by maximum likelihood (counting co-occurrences in training data) results in the well known na\"ive Bayes probability estimator. This reduces the variance of the density estimator but introduces bias whenever the na\"ive assumption does not hold exactly.

\begin{figure}[b!]\vspace{-10pt}
\centering
\includegraphics[width=.45\textwidth]{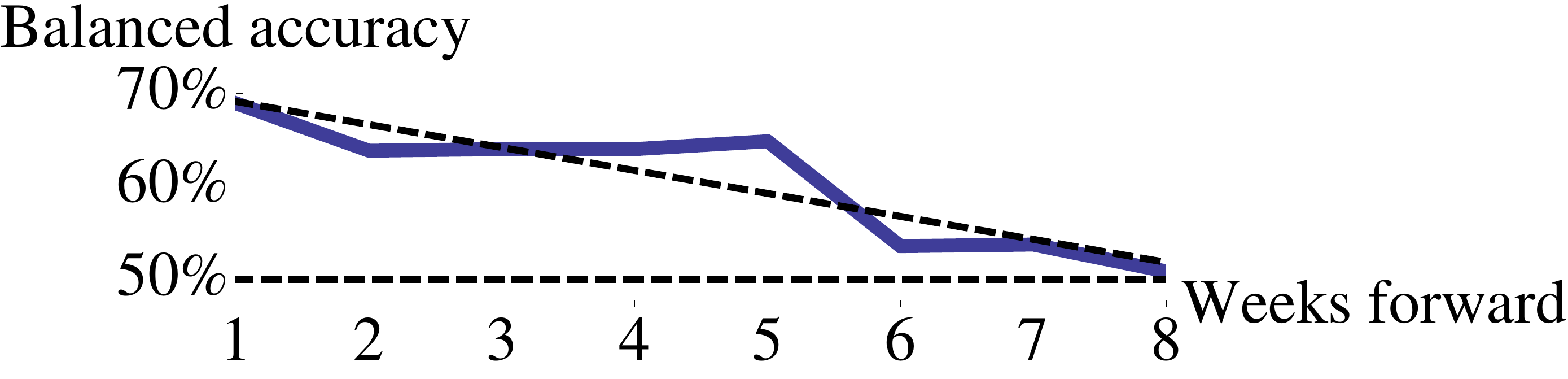}
\caption{In dark blue, the balanced accuracy of predicting cyber attacks against Israel by distance into the future.}
\label{fwdcyber}\end{figure}

To further relieve issues of the sparsity of positives in our data, instead of estimating the conditional probabilities by maximum likelihood, we take a Bayesian approach and assume a Dirichlet prior with symmetric concentration parameter $\alpha$. The prior is our probabilistic belief about the value of these conditional probabilities in the absence of data. Being the conjugate prior to the categorical distribution, a Dirichlet law will also be the distribution for the posterior distribution. Estimating the parameters by maximum a posteriori likelihood is then equivalent to estimating the probabilities by counting frequencies from the data and padding these counts each by $\alpha$. We estimate the marginal $\mathbb P\prns{T_i=1}$ using maximum likelihood because that estimation is not plagued by sparsity.

Estimating the conditional probabilities thus is then done simply by counting data which can be done exceedingly fast. For each event, source, and look-back $k$, one simply counts occurrences in bins of the feature and keeps such tallies separately for when $k+1$ weeks forward (only if within the training data) has been positive or negative. The computation involved in this procedure scales as the product of the length of the data and the number of features.

To make a prediction we check whether $\text{\eqref{density}}\geq p^*$ for a chosen threshold $p^*$. Due to the bias introduced by making the na\"ive assumption, we select $p^*$ by four-fold cross-validation on the training data (up to March 2012) to maximize balanced accuracy.

\subsection{Results}
\begin{table}[t!]
\centering
\begin{tabular}{ll|ll}
Targets & BAC & Perpetrators & BAC\\\hline
Israel & 68.9\% & Anonymous & 70.3\%\\
Germany & 65.4\% & AnonGhost & 70.8\%\\
South Korea & 63.1\% & LulzSec & 60.6\%\\
United Kingdom & 65.5\% & Guccifer & 66.7\%
\end{tabular}
\caption{Accuracy of predicting cyber attacks against and by a selection of entities.}
\label{cybertable}\vspace{-10pt}\end{table}

We first apply this method to predicting cyber campaigns against Israel. For reference, most of these are perpetrated by groups AnonGhost and Anonymous under the banner of \#OpIsrael. Testing on April 2012 to July 2013, we get a true positive rate of \textbf{70.0\%} and a true negative rate of \textbf{67.8\%} giving a balanced accuracy of \textbf{68.9\%}. Our accuracy diminishes as we try to predict farther into the future, as depicted in Figure \ref{fwdcyber}.

By inspecting the trained conditional probabilities we can see which were the most impactful features to sway our belief one way or the other. In this case, swaying our belief most toward predicting positively were if many blog and mainstream mentions of cyber attack appeared in recent weeks and if many social media mentions of protest in Israel appeared in recent weeks as well as forward-looking mentions on social media of a protest in the week to come.

We apply the same method to predicting attacks against three other country entities and to predicting campaigns perpetrated by a selection of four hacktivist groups. The results are reported in Table \ref{cybertable}.

Where hacktivism campaigns are often reactions to developments that do not necessarily at first involve the hacktivist organization, incorporating in some way event mentions involving other entities could boost performance. However, it is not immediately clear how to do so without introducing too many redundant and obfuscating features that will result in overfitting and poor out-of-sample accuracy. Using abstractions as in \cite{radinsky} is one possible way to improve this method.

\end{document}